\documentclass[aps,pre,floats,twocolumn,showpacs]{revtex4-1}
\usepackage{amsmath}                       
\usepackage{amsfonts}
\usepackage{amssymb}
\usepackage{wasysym}
\usepackage{longtable}

\usepackage{graphicx,epsfig}
\usepackage{graphics,dcolumn,bm,fleqn,epic,eepic,float}
\usepackage{amssymb,amsmath,multirow,rotate,color}
\usepackage{subfigure}
\usepackage{color}

\newcommand{\R}{\mathbb{R}}

\begin{document}

\title{Controlling centrality in complex networks}

\author{V. Nicosia$^{1,2}$, R. Criado$^3$, M. Romance$^3$,
  G. Russo$^{2,4}$, V. Latora$^{1,2}$\footnote{Corresponding author:
    latora@ct.infn.it}}

\affiliation{%
  $^1$Dipartimento di Fisica ed Astronomia, Universit\`a
  di Catania, and INFN Sezione di Catania, 95123 Catania, Italy.}
\affiliation{%
  $^2$Laboratorio sui Sistemi Complessi, Scuola Superiore di
  Catania, Universit\`a di Catania, 95123 Catania, Italy.}
\affiliation{%
  $^4$Departamento de Matem\'atica Aplicada, Universidad Rey
  Juan Carlos, 28933 Madrid, Spain.}  
\affiliation{%
  $^4$Dipartimento di Matematica e Informatica, Universit\`a 
  di Catania, Via S. Sofia, 64, 95123 Catania, Italy}

\begin{abstract}
  Spectral centrality measures allow to identify influential
  individuals in social groups, to rank Web pages by their popularity,
  and even to determine the impact of scientific researches.  The
  centrality score of a node within a network crucially depends on the
  entire pattern of connections, so that the usual approach is to
  compute the node centralities once the network structure is
  assigned. We face here with the inverse problem, that is, we study
  how to modify the centrality scores of the nodes by acting on the
  structure of a given network.  We prove that there exist particular
  subsets of nodes, called controlling sets, which can assign
  any prescribed set of centrality values to all the nodes of a graph,
  by cooperatively tuning the weights of their out-going links. We
  show that many large networks from the real world have surprisingly
  small controlling sets, containing even less than $5-10\%$ of the
  nodes. These results suggest that rankings obtained from spectral
  centrality measures have to be considered with extreme care, since
  they can be easily controlled and even manipulated by a small group
  of nodes acting in a coordinate way.
\end{abstract}

\maketitle

Modelling social, biological and information-technology systems as
complex networks has proven to be a successful approach to understand
their function \cite{bocca,arenas,bbv,fortunato}.  Among the various
aspects of networks which have been investigated so far, the issue of
\textit{centrality}, and the related problem of identifying the
central elements in a network, has remained pivotal since its first
introduction. The idea of centrality was initially proposed in the
context of social systems, where it was assumed a relation between the
location of an individual in the network and its influence and power
in group processes \cite{bavelas,wasserman94}. Since then, various
{\em centrality measures} have been introduced over the years to rank
the nodes of a graph according to their topological
importance. Centrality has found many applications in social systems
\cite{wasserman94}, in biology \cite{jmbo01} and in man-made
spatial networks \cite{ba00,porta,spatial}.

Among the various measures of centrality, such as those based on
counting the first neighbours of a node (degree centrality), or the
number of shortest paths passing through a node (betweenness
centrality)~\cite{Freeman79,barthelemy}, a particularly important
class of measures are those based on the spectral properties of the
graph \cite{PF_spectral}.  Spectral centrality measures include the
{\em eigenvector centrality} \cite{Bon1,Bon3}, the {\em alpha
  centrality} \cite{Bon2}, {\em Katz's centrality} \cite{Katz} and
{\em PageRank} \cite{BrinPage}, and are often associated to simple
dynamics taking place over the network, such as various kinds of
random walks \cite{delvenne,delosrios,randomwalks}.  As representative
of the class of spectral centralities, we focus here on 
eigenvector centrality, which is based on the idea that the importance
of a node is recursively related to the importance of the nodes
pointing to it.  Given an unweighted directed graph $G=( V, E)$ with
$N=|V|$ nodes and $K=|E|$ links, described by the $N\times N$
adjacency matrix $A$, the eigenvector centrality $\mathbf{c_0}$ of $G$
is defined as the eigenvector of $A^{t}$ associated to the largest
eigenvalue $\rho_0$, which in formula reads $A^t\mathbf{c_0} =\rho_0
\mathbf{c_0}$ \cite{Bon1,Bon2,Bon3}. If the graph is strongly
connected, then the Perron-Frobenius theorem guarantees that
$\mathbf{c_0}$ is unique and positive. Therefore, $\mathbf{c_0}$ can
be normalised such that the sum of the components equals 1, and the
value of the $i$-th component represents the centrality score of node
$i$, i.e. the fraction of the total centrality associated to node
$i$. In this Article we show how to change the eigenvector centrality
scores of all the nodes of a graph by performing only local changes 
at node level. As a first step (see the Methods Section) we have
proved that, given any arbitrary positive vector $\mathbf{c}\in \R^N$,
$\mathbf{c}>0$, and $\mathbf{c} \neq \mathbf{c_0}$, it is
\textit{always} possible to assign \textit{the weights of all the
  links} of a strongly--connected graph $G$ and to construct a new
weighted network $G_\omega$, with the same topology as $G$ and with
eigenvector centrality equal to $\mathbf{c}$:
\begin{equation}
\label{eq1}
  A^t_\omega \mathbf{c}=\rho \mathbf{c}, 
\end{equation}
where $A_{\omega}$ is the weighted adjacency matrix of $G_\omega$. 

\begin{figure*}
\centering
\includegraphics[width=2.1in]{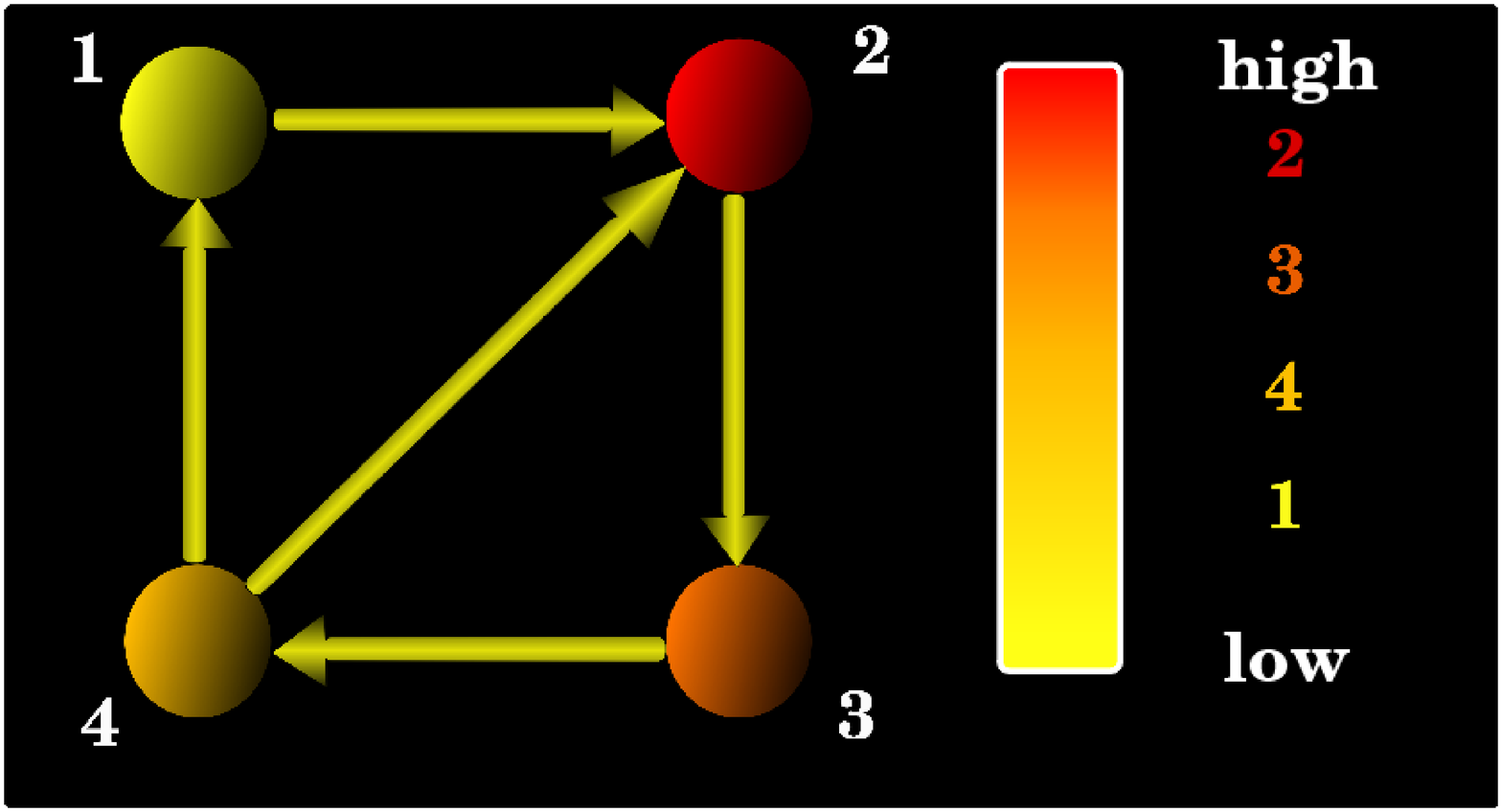}
\includegraphics[width=2.1in]{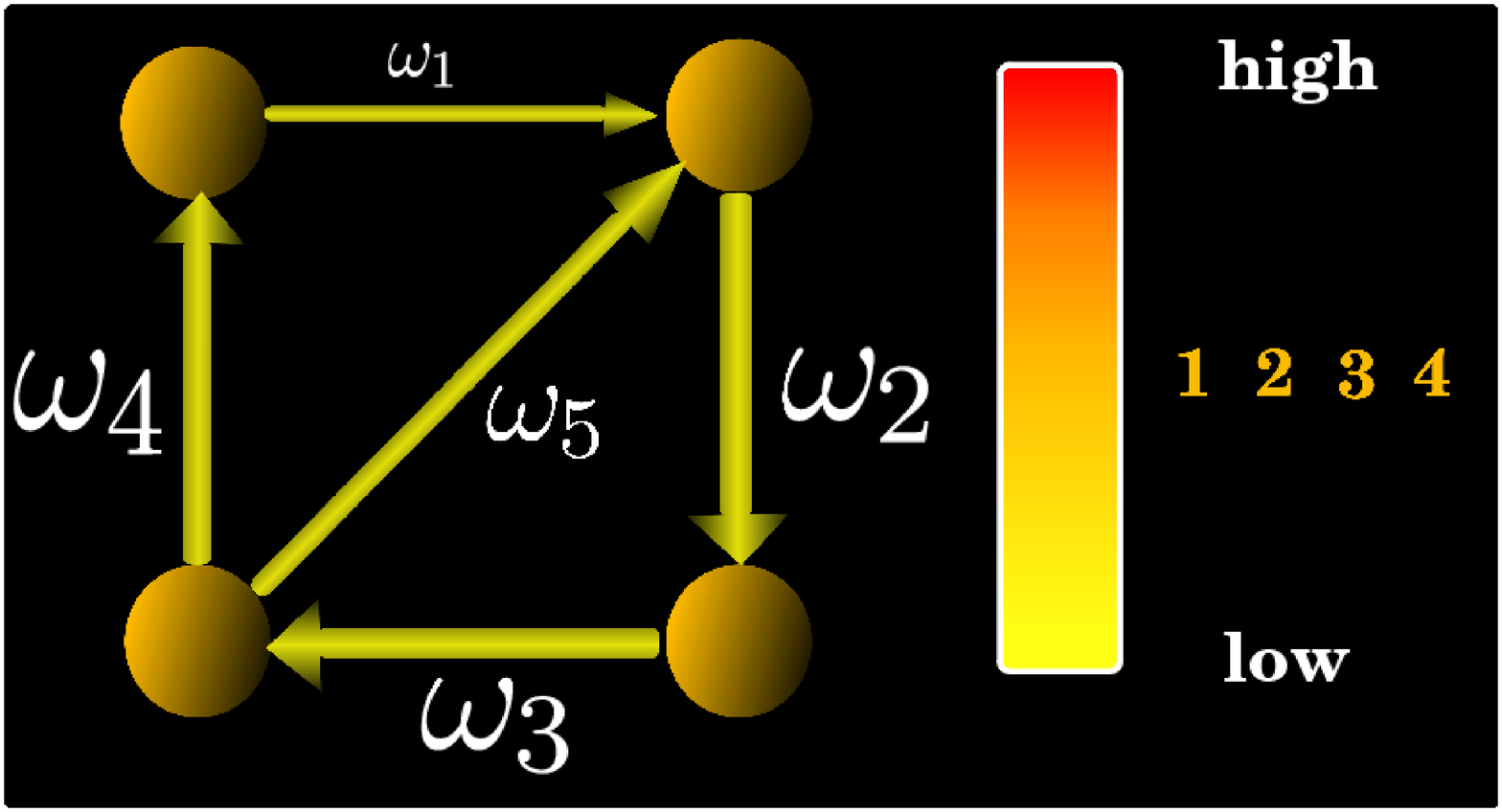}
\includegraphics[width=2.1in]{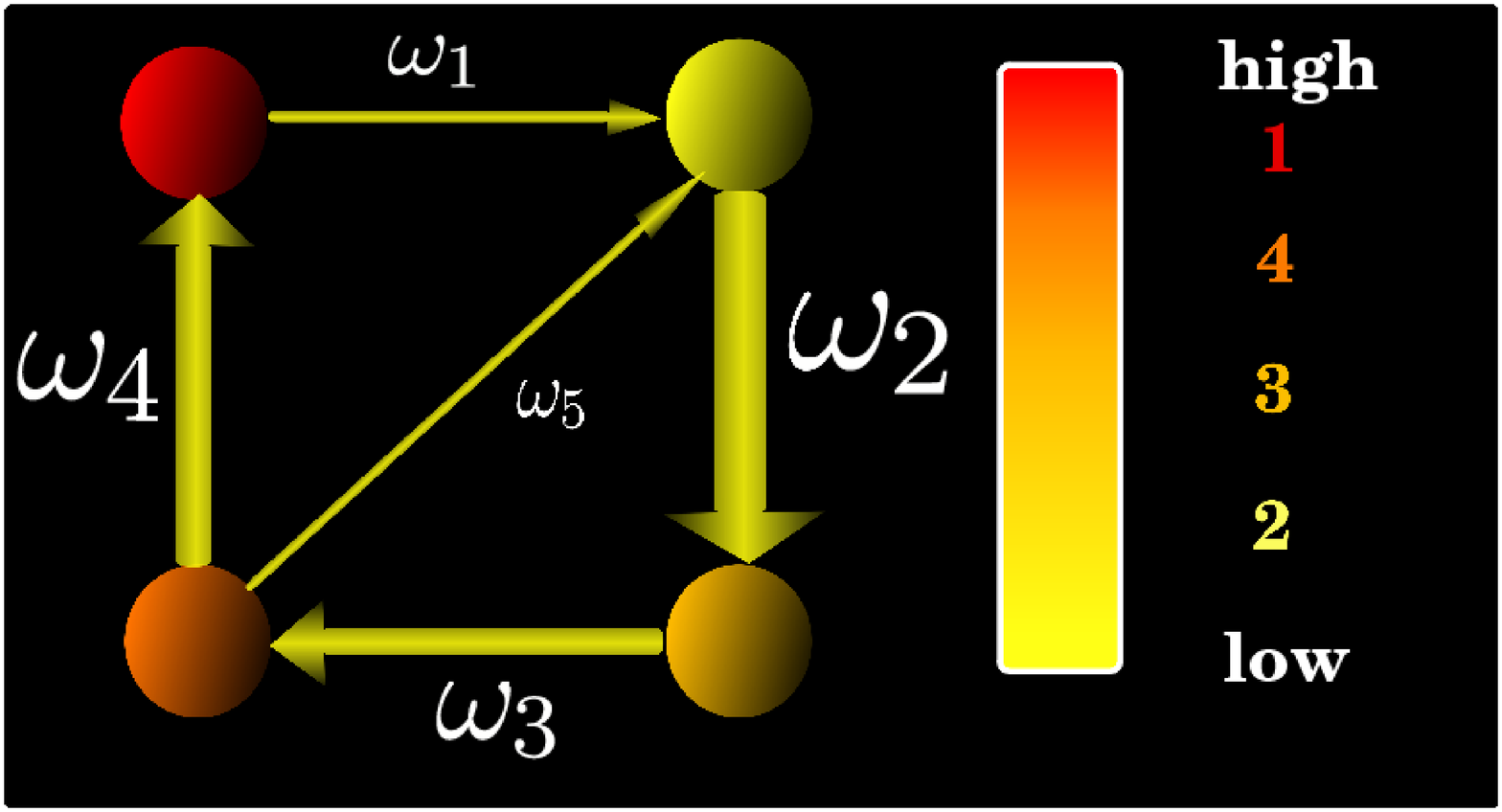}
\caption{{\bf An example of how to tune the link weights to change the
    node centrality scores.}  The graph $G$ with $N=4$ nodes and $K=5$
  links shown in panel (a) is strongly-connected and has an
  eigenvector centrality $\mathbf{c}_0 = \{0.18, 0.33, 0.27, 0.22\}$.
  By ranking the nodes according to the components of $\mathbf{c}_0$,
  we obtain that node $2$ is the most central one, followed in order
  by node $3$, node $4$, and node $1$. We can now set the weights of
  the five links $\bm{\omega} = \{\omega_1, \omega_2, \omega_3,
  \omega_4, \omega_5 \}$ in such a way that Eq.\ref{eq1} is satisfied
  for any given centrality vector $\mathbf{c} \neq \mathbf{c_0}$.  For
  instance, we can get a weighted network $G_\omega$ in which all
  nodes have the same centrality, by solving Eq.\ref{eq1} with a
  centrality vector $\mathbf{c} = \{1/4, 1/4, 1/4, 1/4\}$ and $\rho
  =3.0$. We obtain a vector of weights: $\bm{\omega} = \{\alpha, 3, 3,
  3, 3-\alpha \}$ which, for $0<\alpha<3$, guarantees that all the
  link weights of the graph are positive. The resulting network
  $G_\omega$ is shown in panel (b).  As expected, we have $K-N=1$ free
  parameter (namely $\alpha$) since the graph has $N=4$ nodes and
  $K=5$ links. Instead, if we want to reverse the original node
  ranking we can solve the system $A^{t}_{\omega} \mathbf{c} = \rho
  \mathbf{c}$ with a centrality vector $\mathbf{c}=\{0.5, 0.05, 0.2,
  0.25\}$. Notice that, in this case, the ranking induced by
  $\mathbf{c}$ is exactly the opposite of the one induced by
  $\mathbf{c_0}$: now node 1 is the most central one, followed in
  order by node 4, node 3, and node 2. The solution of
  Eq.\ref{eq:sistema} gives $\bm{\omega} = \{\alpha, 12, 15/4, 6,
  (3-10\alpha)/5\}$, corresponding to a weighted network $G_\omega$
  with all positive weights whenever $0<\alpha <3/10$. The resulting
  network is shown in panel (c).}
\label{fig:smallgraph}
\end{figure*}

This is illustrated in Fig.~\ref{fig:smallgraph} for a graph with
$N=4$ nodes and $K=5$ links. In the original unweighted graph $G$,
node $2$ is the node with the highest eigenvector centrality, followed
in order by node $3$, node $4$, and node $1$. Now, if we have the
possibility of tuning the weights of each of the five links, we can
set any centrality value to the nodes of the graph. In figure we show,
for instance, how to fix the weights of the five links in order to
construct: {\em i)} a weighted network $G_\omega$ in which all nodes
have the same centrality score, and {\em ii)} even a weighted network
$G_\omega$ in which the centrality ranking is totally reversed with
respect to the ranking in $G$.

As shown in the example, given a graph $G$, by controlling the weights
of all the links, it is always possible to set any arbitrary vector
$\mathbf{c}$ as the eigenvector centrality of the graph. However,
tuning the weights of all the $K$ links of a given network is
practically unfeasible, especially in large systems. Fortunately, this
is not necessary, either. In fact, in the case of
Fig.~\ref{fig:smallgraph}, a weighted graph with all nodes having the
same centrality score can also be obtained by changing the weights of
only four links, while leaving unchanged the weight of the link from
node 1 to node 2. More in general, it can be proved that the
eigenvector centrality of the whole network can be controlled by
appropriately tuning the weights of just $N$ of the $K$ links.  The
only constraint is that the $N$ links must belong to a subset $E'
\subseteq E$ such that, for every node $i\in V$, there is a link
$\ell\in E'$ {\sl pointing~to}~$i$ (see Methods Section).

\begin{figure*}
\centering
\includegraphics[width=6.2in]{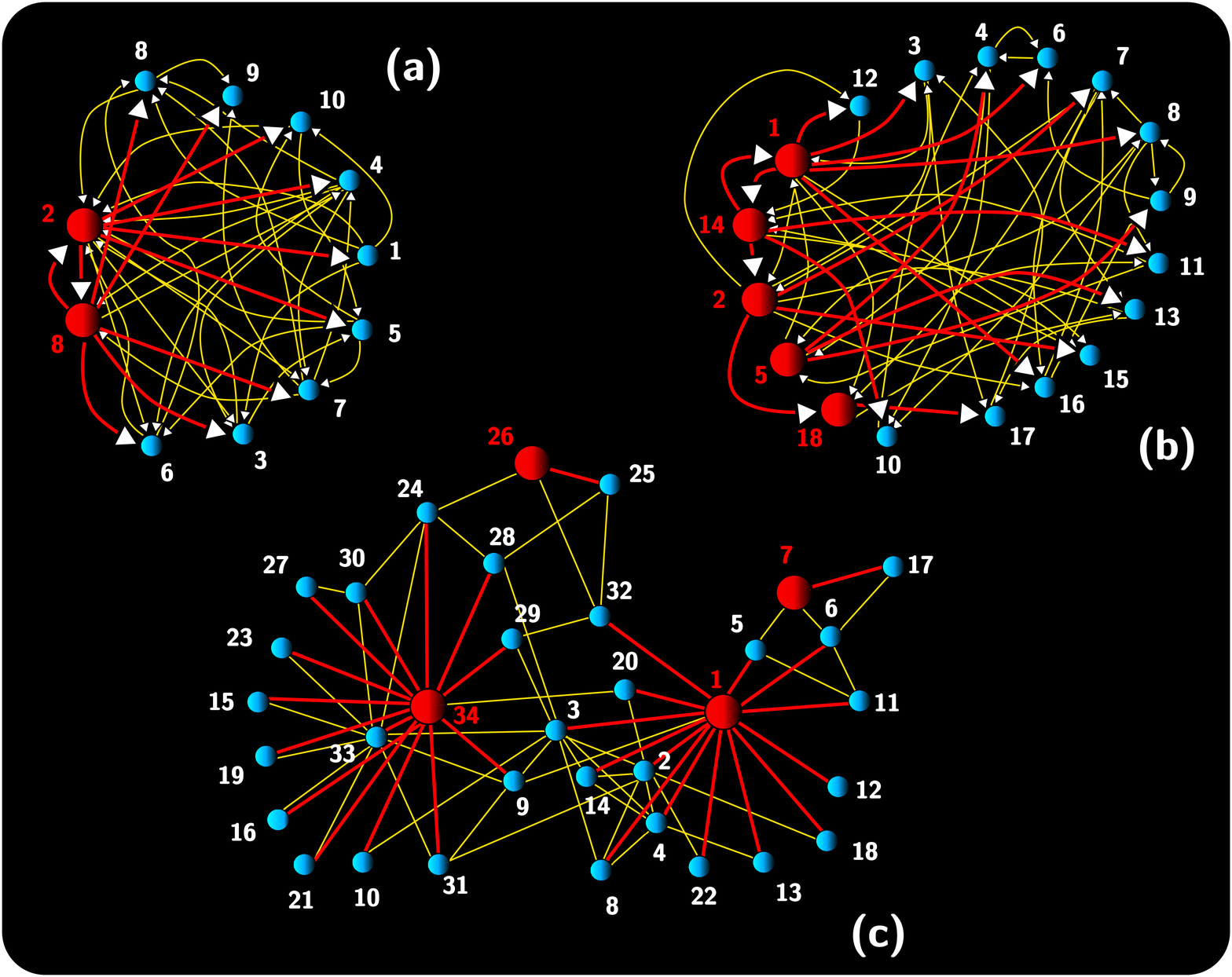}
\caption{{\bf Minimum controlling sets in three real social networks.}
  The graph in panel (a), with $N=11$ vertices and $K=41$ arcs,
  shows who asks who for an opinion among the members of the
  student government of the University of Ljubljana in 1992
  \cite{hlebec}.  The minimum controlling set is made by the two
  nodes marked in red, namely node 2 and node 8. These two nodes are
  linked to each other and point to all the remaining nodes in the
  graph. Therefore, nodes 2 and 8, by cooperatively modifying the
  weights of their red links, can set any arbitrary eigenvector
  centrality to the entire system. The graph in panel (b) has $N=18$
  nodes and $K=55$ arcs, and describes the social relations between
  the monks of an isolated contemporary American monastery, as
  recorded by Sampson in 1969 \cite{sampson}. Here, the minimum
  controlling set contains five nodes, shown in red.  In this case,
  the subset of links $E'$ (red links), does not contain links
  pointing to node 5, so that the red nodes can control the centrality
  of all the network nodes, except node 5. Finally, the Zachary's
  karate club network shown in panel (c) has $N=34$ nodes and $K=78$
  undirected edges, and describes the social network of friendships
  among the members of a US university karate club in 1970
  \cite{zachary}.  In this network, the minimum controlling set
  contains node 1, the instructor Mr. Hi, node 34, the club
  administrator Mr. John A, and also nodes 7 and 26. Notice that just two 
nodes, namely 1 and 34, can control the centrality of
  $95\%$ of the graph nodes.}
\label{fig:real_social}
\end{figure*}

This is illustrated in Fig.~\ref{fig:real_social} for three real
social networks. In each of the three cases, it is possible to set any
arbitrary eigenvector centrality by changing only the weights of the
red arcs, while keeping unchanged (and equal to $1$) the weights of
all remaining arcs, shown in yellow.  The nodes from which the links
in $E'$ originate are also coloured in red, and are referred to as a
{\em controlling set} of the network (see Methods Section).  What is
striking is that, in each of the three networks, the set $E'$ can be
chosen in such a way that all the links in $E'$ originate from a
relatively small subset of nodes. For instance, the controlling set
reported for the student government network of the University of
Ljubljana contains only two nodes.  This is also a {\em minimum
  controlling set}, since the graph does not admit another controlling
set with a smaller number of nodes. This finding indicates that only
two members of the student government, namely node 2 and node 8, can
in principle set the centrality of all the other members by
concurrently modifying the weights of some of their links.  It is in
fact reasonable to assume that the weight of the directed link from
$i$ to $j$, representing in this case the social credit (in terms of
reputation, esteem or leadership acknowledgement) given by individual
$i$ to individual $j$, can be strengthen or decreased {\em only} by
$i$. Consequently, nodes 2 and 8 can modify at their will the weights
of their out-going links and, If these changes are opportunely
coordinated, they can largely alter the actual roles of all the other
individuals.  Analogously, only five monks can control the centrality
of the Sampson's monk network, while only 4 members of the Zachary's
karate club network can set the eigenvector centrality of the
remaining 30 members.

\bigskip A question of practical interest is to investigate the size
${\cal C} \equiv |C^{*}|$ of the minimum controlling set in various
complex networks.  When ${\cal C}$ is small with respect to $N$, then
the centrality of the network is easy to control. Conversely, when the
number of nodes in the minimum controlling set is large, the network
$G$ is more robust with respect to centrality manipulations. We have
used two greedy algorithms to compute approximations of minimum
controlling sets in various real systems (see Methods Section).  In
Table~\ref{table} we report the best approximation for ${\cal C}$,
i.e. the size of the smallest controlling set $\overline{C}$ produced
by either of the two algorithms in networks whose sizes range from
hundreds to millions of nodes.  In the majority of the cases we have
found unexpectedly small controlling sets, containing only up to
$10-20\%$ of the nodes of the network. For instance, in the graph of
Jazz musicians, there exists a controlling set made by just $16$ of
the $198$ musicians. These $16$ individuals alone can, in principle,
decide to set the popularity of all the other musicians, enhancing the
centrality of some of the nodes and decreasing the centrality of
others, just by playing more or less often with some of their first
neighbours. Among all the networks we have considered, the one with
the smallest controlling set is the Wikipedia talk communication
network, a graph with 2,394,385 nodes in which just $2\%$ of nodes are
able to alter the centrality of the entire system. The quantities in
parenthesis indicate that for this network a set of just $1\%$ of the
nodes can control the centrality of $99\%$ of the nodes.

\begin{table*}
  
  
  
  
  \centering
  \begin{tabular}{|l|c|c|c|c|}                          

    \hline                                              
    \textbf{Network} & \textbf{$N$} & \textbf{$\langle k \rangle$} &         
    \textbf{$\mathcal{C}(G)$} &                         
    \textbf{$\mathcal{C}(G^{rnd})$} \\ \hline           
    \hline                                              
    Web~(Berkley and Stanford) \cite{leskovec-lang}          
    & 654782 & 22.2 & 8\% (3\%$\rightarrow$95\%) & 12\%\\\hline     
    Web (Google)~\cite{leskovec-lang}                        
    & 875713 & 11.1 & 15\% (9\%$\rightarrow$94\%) & 22\% \\ \hline  
    Web (NotreDame)~\cite{webnd}                         
    & 325729 & 9.2 & 13\% (8\%$\rightarrow$95\%) & 21\% \\ \hline   
    Web (Stanford)~\cite{leskovec-lang}                      
    & 281904& 16.4 & 8\% (3\%$\rightarrow$95\%) & 15\%\\\hline      
    \hline                                              
    Jazz musicians~\cite{danonjazz}                     
    & 198 & 27.7 & 8\% (5\%$\rightarrow$97\%)& 13\%\\ \hline        
    Movie actors~\cite{actors}                                
    & 392340 & 7.2 & 11\% (8\%$\rightarrow$97\%)& 22\% \\ \hline    
    Cond-Mat coauthorship~\cite{Newman2001}              
    & 12722 & 6.3 & 23\%(18\%$\rightarrow$93\%) & 29\% \\ \hline    
    AstroPh coauthorship~\cite{Newman2001}               
    & 13259 & 18.7 & 16\% (10\%$\rightarrow$94\%) & 27\% \\ \hline  
    Networks coauthorship~\cite{newman-net}             
    & 379 & 4.8 & 20\% (15\%$\rightarrow$94\%) & 29\%\\ \hline      
    URV email~\cite{urv-email}
    & 1133 & 9.6 & 23\% (16\%$\rightarrow$91\%) & 27\% \\ \hline    
    ENRON email~\cite{leskovec-lang,enron}
    & 2351 & 118.7 & 7\%(4\%$\rightarrow$97\%) & 8\% \\ \hline      
    Email EU-All~\cite{leskovec-kleinberg}                      
    &265214 & 3.5& 16\% (1\%$\rightarrow$85\%)& 26\% \\\hline       
    Wiki-talk~\cite{leskovec-hutt}
    &2394385 & 4.20 & 2\% (1\%$\rightarrow$99\%) & 23\% \\\hline    
    \hline                                              
    Hep-Ph citation~\cite{leskovec-kleinberg}
    & 34401 & 12.25 & 16\% (10\%$\rightarrow$94\%) & 22\%\\ \hline  
    Hep-Th citation~\cite{leskovec-kleinberg}
    & 27400 & 12.7 & 17\%(8\%$\rightarrow$91\%) & 22\%\\ \hline     
    Patents~\cite{leskovec-kleinberg}  
    & 3774768 & 8.75 & 50\% (16\%$\rightarrow$60\%) & 26\% \\\hline
    \hline
    Internet AS~\cite{InternetAS}
    & 11174 & 4.2 & 9\% (8\%$\rightarrow$99\%) & 22\%\\ \hline
    US Airports ~\cite{USAirports}
    & 500 & 11.9 & 14\% (12\%$\rightarrow$97\%) & 19\%\\ \hline     
    US Power Grid \cite{watts}                           
    & 4941 & 5.33 & 33\% (29\%$\rightarrow$95\%)& 23\%\\ \hline     
    roadnet CA~\cite{leskovec-lang}
    & 1965206 & 5.63 & 31\% (30\%$\rightarrow$96\%)& 23\%\\ \hline           
    roadnet PA~\cite{leskovec-lang}
    & 1088092 & 5.67 & 33\% (30\%$\rightarrow$967\%) & 23\% \\ \hline         
    roadnet TX~\cite{leskovec-lang}
    & 1379917 & 5.57  & 33\% (30\%$\rightarrow$97\%) & 23\%  \\ \hline 
    Electronic circuit (s208 st)~\cite{Alon}     
    & 123         & 3.1    & 29\% (26\%$\rightarrow$96\%) & 28\%  \\ \hline     
    Electronic circuit (s420 st)~\cite{Alon}
    & 253         & 3.1    & 29\% (25\%$\rightarrow$96\%) & 28\%  \\ \hline
    Electronic circuit (s838 st)~\cite{Alon}        
    & 513         & 3.2    & 29\% (25\%$\rightarrow$96\%) & 28\%  \\ \hline
    \hline                                              
    Wordnet~\cite{wordnet}
    & 77595 & 3.44 & 26\% (19\%$\rightarrow$92\%)& 26\%\\ \hline    
    USF Words associations~\cite{usf-norms}
    & 10618 & 13.6 & 22\% (8\%$\rightarrow$56\%) & 25\%\\ \hline    
    \hline                                              
    PGP~\cite{pgp}
    & 10680 & 4.5 & 22\%  (18\%$\rightarrow$77\%) & 29\% \\ \hline  
    Amazon~\cite{amazon}                                
    & 410236 & 16.36 & 17\% (9\%$\rightarrow$91\%)& 17\% \\\hline   
    Epinions~\cite{epinions}                            
    & 75879 & 13.41 & 22\% (19\%$\rightarrow$95\%) & 18\% \\\hline 
    Gnutella~\cite{leskovec-kleinberg,gnutella2}                  
    & 62586 & 4.72 & 19\%(11\%$\rightarrow$62\%) & 31\% \\\hline    
    PolBlogs~\cite{polblogs}                         
    & 1224 & 31.2 & 13\% (8\%$\rightarrow$94\%) & 18\%\\\hline      
    PolBooks~\cite{polbooks}                         
    & 105 & 8.4 & 15\% (12\%$\rightarrow$98\%) & 22\%\\\hline       
    Slashdot~\cite{leskovec-lang}                            
    & 82168 & 23.08 & 25\% (21\%$\rightarrow$58\%)& 27\% \\\hline   
    Wiki-vote~\cite{leskovec-hutt}
    & 8298 & 25.00 & 16\% (15\%$\rightarrow$99\%) & 20\%  \\\hline  
  \end{tabular}                                         
  \caption{ Number of nodes $N$, average degree $\langle k \rangle$,
    and the relative size ${\cal C}(G)$ of the mimimum controlling set
    found in 35 different real world networks. The values of ${\cal
      C}(G)$ reported are expressed as percentage of the network size
    $N$. The algorithms used to find approximations of minimum
    controlling sets, mark as \textit{controllers} also nodes not
    controlling other nodes, simply because, at a certain iteration of
    the greedy procedure, they have remained with no out-going
    links. Therefore, we also report in parenthesis the relative size
    of the \textit{effective} minimum controlling set and the
    percentage of the controlled nodes.  The notation $x \% \rightarrow
    y \%$, indicates that $x \%$ of the nodes is able to control the
    centrality of $y \%$ of the network.  In the rightmost column we
    report, for each network, the relative size of the controlling set
    in randomized versions which preserve the original
    degree sequence. We have considered averages over 100
    different randomizations. From top to bottom, the networks are
    divided into six classes, respectively World-Wide-Web,
    collaboration/communication, citation, spatial, words and
    socio--economical networks.}
  \label{table}                                                
\end{table*}    

\begin{figure}
\centering
\includegraphics[scale=0.3]{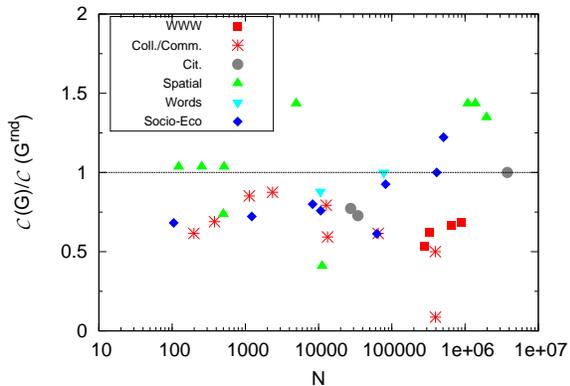}
\caption{{\bf Relative size of the minimum controlling set in various
    real systems.}  We report, as a function of $N$, the ratio between
  the sizes of the minimum controlling set in real networks and in
  their respective randomized versions (we have considered averages
  over 100 different realizations). Different symbols and colors refer
  to the six network classes considered in Table~\ref{table}. The
  observed ratio is lower than 1 in most of the cases, with the
  smallest values corresponding to collaboration/communication
  systems, WWW and socio-economical networks. The ratio is equal to 1
  in three cases.  The networks with ratio larger than 1, with the
  exception of one socio-economical system (namely Epinions), are all
  spatially constrained systems: three electronic circuits, the US
  power grid, and three road networks.  }
\label{fig:Crand}
\end{figure}

For each real network $G$, we have also computed the typical size
$\mathcal{C}(G^{rnd})$ of the minimum controlling set in its
randomised counterpart (see the rightmost column in
Table~\ref{table}).  In particular, we have considered a randomisation
which preserves the degree sequence of the original graph. In most of
the cases $\mathcal{C}(G) \le \mathcal{C}(G^{rnd})$, relevant
exceptions being some spatial man-made networks, such as power grids,
road networks and electronic circuits, and also the patents citation
network. This fact suggests that, in the absence of other limitations,
such as strong spatial/geographic constraints~\cite{spatial}, the
structure of real networks has naturally evolved to favour the control
of spectral centrality by a small group of nodes.  To better compare
the controllability of networks with different sizes, we report in
Fig.~\ref{fig:Crand} the ratio
$\frac{\mathcal{C}(G)}{\mathcal{C}(G^{rnd})}$ as a function of the
number of graph nodes $N$. The smallest values of the ratio
$\frac{\mathcal{C}(G)}{\mathcal{C}(G^{rnd})}$ are found for
collaboration/communication systems, WWW and socio-economical
networks. The five most controllable networks are respectively
Wiki-talk, Internet at the AS level, movie actors, the Stanford World
Wide Web, and the collaboration network of researchers in
astrophysics.  These are all networks in which single nodes can tune,
at their will, the weights of their out-going links.  A scientist can
decide whether to weaken or strengthen the connections to some of the
collaborators. The administrators of an Internet Autonomous System can
control the routing of traffic through neighbouring ASs, by modifying
peering agreements \cite{vespignani_book}.  And, similarly, the owner
of a Web page can change the weights of hyperlinks, for instance by
assigning them different sizes, colour, shapes and positions in the
Web page.

In this work, we have shown how a small number of entities, working
cooperatively, can set any arbitrary eigenvector centrality for all
the nodes of a real complex network. It is straightforward to extend
our results to other spectral centralities, such as
$\alpha$-centrality and Katz's centrality. Similar arguments can also
be applied, with some limitations, to PageRank: in this case, the
inverse centrality problem has solutions only for some particular
choices of ${\bf c}$. Such findings suggest that rankings obtained
from centrality measures should be taken into account with extreme
care, since they can be easily controlled and even distorted by a
small group of cooperating nodes. The high controllability of real
networks potentially has large social and commercial impact, given
that centrality measures are nowadays extensively used to identify key
actors, to rank Web pages, and also to assess the value of a
scientific research.

\section{Methods}
 
\subsection{Solution to the inverse centrality problem.} The set of $N$
linear equations with $K$ variable weights,
$\omega_1,\ldots,\omega_K$, in Eq.~\ref{eq1} can be rewritten as a
system of $N$ linear equations with $K$ variables:
\begin{equation}\label{eq:sistema}
  B \bm{\omega} =\rho \mathbf{c},
\end{equation}
where now $B$ is a $N\times K$ matrix of real numbers, and
$\bm{\omega}\equiv \{\omega_1,\ldots,\omega_K\}$. Notice that the
linear system in Eq.~\ref{eq:sistema} has solutions since the rank of
$B$ is $N<K$ (all the equations are separated and each of the
variables, $\omega_1,\ldots,\omega_K$, appears in one equation only),
and the in-degree of all nodes is positive by definition.  Hence,
there always exists $\bm{\omega} \in \mathbb{R}^K$ such that
Eq.~\ref{eq1} is satisfied. It is convenient to rewrite
Eq. \ref{eq:sistema} in a form that emphasises the dependence of
matrix $B$ from $c$. We choose to label the arcs as follows: $(i,l)$,
$l=1\ldots k^{in}_i$ denotes the $l$-th arc entering node $i$, where
$k^{in}_i$ is the in--degree of node $i$.  Likewise, $S_{i,l}$ is the
source of arc $(i,l)$, while $\omega_{i,l}$ is the corresponding
weight. Using this notation, the $i$-th component of
Eq.~\ref{eq:sistema} can be written as:
\begin{equation}
  \sum_{l=1}^{k^{in}_i}\omega_{i,l}c_{S_{i,l}} = \rho c_i
  \label{eq:eq2}
\end{equation}
By direct computation, one positive solution of Eq. \ref{eq:eq2} is
given by 
\begin{equation}
  \omega_{i,1}=\omega_{i,2} = \cdots =\omega_{i,k^{in}_i} = \frac{\rho
    c_i}{\sum_{l=1}^{k^{in}_i} c_{S_{i,l}} }
\end{equation}
where $i=1\ldots N$, and by continuity there are infinite many
solutions such that $\omega_{i,l}$ are all positive. In particular, if
for node $i$ we have $k^{in}_i=1$, then the $i$-th equation of
Eq.~\ref{eq:eq2} has a unique solution, while if $ k^{in}_i>1$, there
are always infinitely many solutions depending on $k^{in}_i -1$
parameters.  Summing up, Eq.~(\ref{eq:sistema}) has only one solution
if all the node in-degrees are equal to one, while there are, in
general, infinitely many solutions depending on $K-N$
parameters. Notice that $\rho$ can be different from $\rho_0$,
meaning that it is also possible to set the value of the largest
eigenvalue of the weighted graph.

\subsection{Tuning a subset of the graph links.}  Here, we show that it is
not necessary to fix the weights of all the graph links in order to
get an arbitrary centrality vector $\mathbf{c}>0$. In fact, given a
subset of links $E'\subseteq E$ containing at least one incoming link
for each node, it is sufficient to assign some positive weights
$\tilde\omega(\ell')$ to each $\ell'\in E'$, while keeping
$\omega(\ell)$ constant $\forall \ell\in E \setminus E'$, for instance
all equal to $1$, such that the resulting weighted graph has
eigenvector centrality equal to $\mathbf{c}$.
Without loss of generality we can assume that the first $k^{c}_i>0$
incoming links of each node $i$ belong to $E'$, so that the components
of Eq.~\ref{eq:eq2} can be written as:

\begin{equation}
  \sum_{l=1}^{k^{c}_i}\omega_{i,l}c_{S_{i,l}} = \rho c_i -
  \sum_{l=k^{c}_i + 1}^{k^{in}_i}\omega_{i,l}c_{S_{i,l}},\quad
  i=1\ldots N
  \label{eq:eq3}
\end{equation}
Therefore, since $c_i>0$ for each $1\le i\le N$, then there is a
$\rho_0>0$ such that for every $\rho>\rho_0$
\begin{equation}
  \rho c_i- \sum_{l=k^{c}_i + 1}^{k^{in}_i}\omega_{i,l}c_{S_{i,l}}>0,
  \quad i=1\ldots N
\end{equation}
and hence, by a similar continuity argument as above, we can ensure
that there are infinitely many positive solutions to Eq.\ref{eq:eq3}.

\subsection{Finding minimum controlling sets.} A \textit{controlling set} of
graph $G$ is any set of nodes $C\subseteq V$ such that:
\begin{equation}
  \displaystyle V = C\cup\left(\bigcup_{i\in C}\{j \in V:\enspace
    e_{ij}\in E\}\right).
  \label{eq:control_set}
\end{equation}
This means that, for each node $j$ in the graph, at least one of the
two following conditions holds: {\em a}) $j \in C$, or {\em b}) $j$ is
pointed by at least one node in $C$. We use $|C|$ to denote the size
of the controlling set, i.e. the number of nodes contained in $C$.
Finding the \textit{minimum controlling set} $C^*$ of a graph $G$,
i.e. a controlling set having minimal size, is equivalent to computing
the so-called {\em domination number} of $G$.  The domination number
problem is a well known NP-hard problem in graph theory \cite{west}.
Therefore, the size of the minimum controlling set can be determined
exactly only for small $N$ graphs as those in
Fig.~\ref{fig:real_social}. To investigate larger graphs we have used
two greedy algorithms.  The first algorithm, called Top--Down
Controller Search (TDCS), works as follows. We initially set
$G_{t=0}=G$. We select the node $i_0$ with the maximum out-degree in
$G_{t=0}$, and mark it as \textit{controller node}. Then, all the
nodes in the out-neighbourhood of $i_0$ are marked as
\textit{controlled} and are removed from $G_{t=0}$, together with
$i_0$ itself. In this way, we obtain a new graph $G_{t=1}$, and we
store the controller node $i_0$, together with the list of nodes
controlled by $i_0$. Notice that, removing a generic node $j$ from
$G_{t=0}$, also implies that $G_{t=1}$ does not contain any of the
links pointing to $j$ or originating from it.  The same procedure is
iteratively applied to $G_{t=1}$, $G_{t=2}$ and so on, until all the
nodes of $G$ are either marked as controller or as controlled
nodes. The algorithm produces a set $\overline{C} = \{i_0, i_1,
i_2,\ldots\}$, with $|\overline{C}|\ge |C^*|$, which is a controlling
set of $G$ by construction.
The second algorithm is called Bottom--Up Controller Search (BUCS),
and it works as follows. We set $G_{t=0}=G$ and consider the set
$M(0)$ containing all the nodes in $G_{t=0}$ with minimum
in--degree. For each node $i\in M(0)$, we consider the set of nodes
pointing to $i$ and select from this set the node $m_i$ with the
maximal out--degree. This node is marked as \textit{controller}. Then
we obtain a new graph $G_{t=1}$ by removing from $G_{t=0}$ all the
controller nodes $m_i$ for all $i\in M(0)$, together with all the
nodes, marked as \textit{controlled}, pointed by them. The same
procedure is iteratively applied to $G_{t=1}$, $G_{t=2}$ and so on,
until all the nodes of $G$ are either marked as controller or as
controlled nodes. If a graph $G_{t}$ contains isolated nodes, these
are marked as \textit{controller} and removed from $G_{t}$. The
algorithm finally produces a set $\overline{C} = \{i_0, i_1,
i_2,\ldots\}$ which is a controlling set of $G$ by construction.  We
have verified that the controlling sets obtained by both TDCS and BUCS
for each of the networks considered are much smaller than those
obtained by randomly selecting the controlling nodes. Moreover, the
set of controller nodes found by TDCS is in general different from
that obtained on the same network by BUCS.  Also the sizes of the two
controlling sets obtained by the two algorithms are different.  In
particular, we have noticed that in assortative (disassortative)
networks the controlling set produced by TDCS is smaller (larger) than
that produced by BUCS.





\end{document}